\documentclass{article}

\usepackage{PRIMEarxiv}

\usepackage[utf8]{inputenc} 
\usepackage[T1]{fontenc}    
\usepackage{hyperref}       
\usepackage{url}            
\usepackage{booktabs}       
\usepackage{amsfonts}       
\usepackage{nicefrac}       
\usepackage{microtype}      
\usepackage{lipsum}
\usepackage{fancyhdr}       
\usepackage{graphicx}       
\graphicspath{{media/}}     

\title{Deep Anomaly Generation: An Image Translation Approach of Synthesizing Abnormal Banded Chromosome Images
}

\author{
Lukas Uzolas\thanks{Equal contribution}, Javier Rico\textsuperscript{*}

 \\
  Pázmány Péter Catholic University, Budapest, Hungary\\
Autonomous University of Madrid, Madrid, Spain \\
University of Bordeaux, Bordeaux, France\\
  \texttt{lukas@uzolas.com, jvirico@gmail.com} \\
   \And
  Pierrick Coupé \\
  CNRS, Univ. Bordeaux, Bordeaux INP, LaBRI, UMR5800, PICTURA, F-33400\\ Talence, France\\
     \And
  Juan C. SanMiguel \\
  Autonomous University of Madrid \\
  Madrid, Spain \\
     \And
  György Cserey \\
  Pázmány Péter Catholic University \\
  Budapest, Hungary \\
}

\begin{document}
\maketitle

\begin{abstract}
Advances in deep-learning-based pipelines have led to breakthroughs in a variety of microscopy image diagnostics. However, a sufficiently big training data set is usually difficult to obtain due to high annotation costs. In the case of banded chromosome images, the creation of big enough libraries is difficult for multiple pathologies due to the rarity of certain genetic disorders. Generative Adversarial Networks (GANs) have proven to be effective in generating synthetic images and extending training data sets. In our work, we implement a conditional adversarial network that allows generation of realistic single chromosome images following user-defined banding patterns. To this end, an image-to-image translation approach based on self-generated 2D chromosome segmentation label maps is used. Our validation shows promising results when synthesizing chromosomes with seen as well as unseen banding patterns. We believe that this approach can be exploited for data augmentation of chromosome data sets with structural abnormalities. Therefore, the proposed method could help to tackle medical image analysis problems such as data simulation, segmentation, detection, or classification in the field of cytogenetics.

\end{abstract}


\section{Introduction}
In the field of cytogenetics, a discipline that studies the structure and function of chromosomes, karyotyping analysis is an important diagnostic procedure \cite{graham1994automatic}. Using staining techniques, Giemsa staining being the most widely used \cite{Holmquist1982}, the 23 pairs of metaphase chromosomes that determine the chromosome complement of an individual are divided by stripes known as banding patterns, or G-bands when Giemsa stained. Those bands, that serve as landmarks on each healthy chromosome type, are standardized in diagrams, called ideograms. Hence all bands present in an ideogram are present in a properly stained healthy chromosome, while the opposite is not necessarily true \cite{shaffer2016iscn}. Chromosomal abnormalities are the direct cause of genetic diseases \cite{natarajan2002chromosome}, existing in two main types, which are manifested in chromosome number and chromosome structure, the former can be identified by simply counting \cite{gersen1999principles}. 
In contrast, the latter is caused by a small portion of the chromatid arm of a chromosome breaking, recombining, or missing, hence reflecting a deviation from the standardized ideogram, and their detection is an important problem in cytogenetics \cite{li2020cs}.

Extensive works are being undertaken to study how machine learning can help pathology diagnosis. Li et al. \cite{li2020cs}, for example, developed the CS-GANomally for chromosome anomaly detection. However, the lack of data is often mentioned \cite{wu2020new}, especially for diseases with complex karyotypes involving multiple structural abnormalities (e.g. leukemia \cite{Nagy2018,salah2019machine}).

Many applications adopt Generative Adversarial Networks (GANs) as a Data Augmentation (DA) technique to improve the performance of deep learning models \cite{bhattacharya2020gan,xu2020automatic}. This has become a trend in medical imaging, where plenty of examples exist, lung nodules \cite{chuquicusma2018fool}, liver lesions \cite{frid2018gan}, and COVID-19 CT scans \cite{Loey2020}. Wu et al. \cite{wu2018end} proposed MD-GAN to generate various data modes of healthy chromosomes to train a classifier for karyotyping. Furthermore, approaches can be found for DA of images presenting user-defined anomalies. C. Han et al. \cite{Han2019}, for example, used a conditional PGGAN-based DA to generate Magnetic Resonance (MR) images with brain metastases at desired positions, improving disease detection \cite{karras2017progressive}. Another approach of generating data for a desired acquisition modality is through image-translation networks, such as pix2pix \cite{isola2017image} or CycleGAN \cite{zhu2017unpaired}. For example, Choi and Lee \cite{choi2018generation} used pix2pix to generate structural MR images from Positron Emission Tomography images, and Yan et al. \cite{yan2019domain} used CycleGAN for domain adaptation of MR images.

While MD-GAN is used to generate healthy chromosomes \cite{wu2018end}, and CS-GANomaly is employed for detection of abnormal ones \cite{li2020cs}, we attempt to generate abnormal chromosomes through specific control over the banding patterns. To the best of our knowledge, this has not been attempted before. To achieve this, we condition a pix2pix network with 2D chromosome banded segmentation masks to synthesize realistic single chromosome images following user-defined banding patterns, allowing for the simulation of chromosomes with abnormalities of the structural type.

\begin{figure}[t]
\centering
\includegraphics[width=0.8\textwidth]{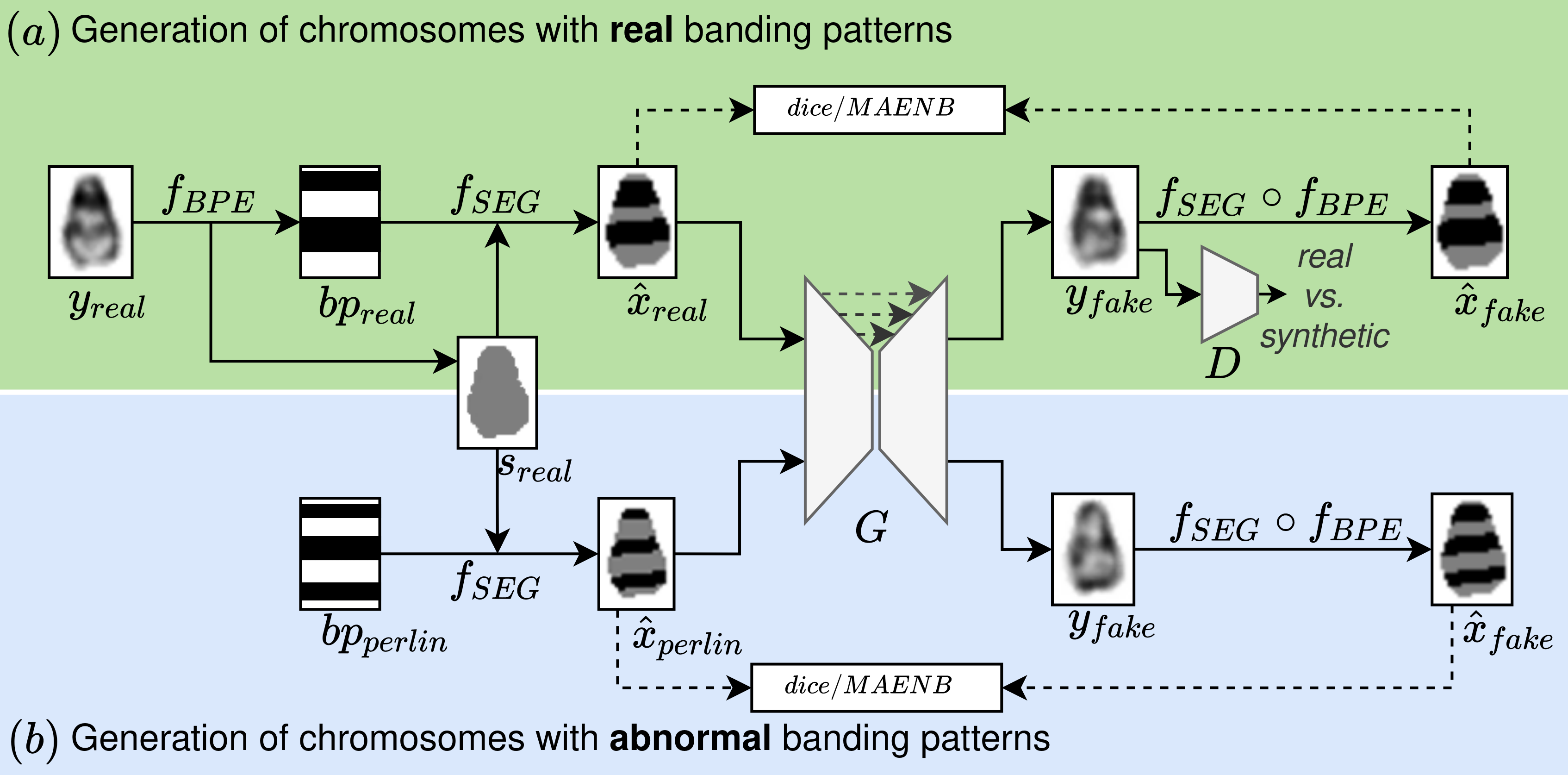}
\caption{Overview of the proposed method. (1) Chromosome generation based on extracted banding patterns from real chromosomes. (2) Like in (1) but with randomly generated Perlin noise banding patterns, resulting in abnormal chromosomes.}

\label{fig:method_overview}
\end{figure}

\section{Method} \label{section:method} 
We use pix2pix \cite{isola2017image} for conditionally generating chromosome images based on a user-defined banding pattern and a shape mask. We extract 1D binary banding patterns in an automatic fashion, which are back-projected onto chromosome shape images, to finally generate 2D banded segmentation masks. The chromosome images, in combination with their respective 2D banded segmentation masks, can be used to train a pix2pix network (see Section \ref{section:pix2pix}). After training, realistic-looking chromosomes can be generated that display a desired band configuration (see Section \ref{section:abnormal_chromosomes}). An overview of this approach can be seen in Figure \ref{fig:method_overview}. More details are given in the following sections. 

\subsection{Real Banding Pattern-Based Chromosome Generation}
\paragraph{Banding Pattern Extraction.} \label{section:banding_pattern_extraction}


An automated banding pattern extraction procedure is implemented, as the corresponding ideograms can not be used as banding pattern approximations. This is due to several reasons. Firstly, the lack of annotations, secondly, the variable regions of healthy chromosomes \cite{shaffer2016iscn}, and thirdly, the presence of unhealthy chromosomes in the data set.



\begin{figure}[t]
    \centering
    \includegraphics[width=0.68\textwidth]{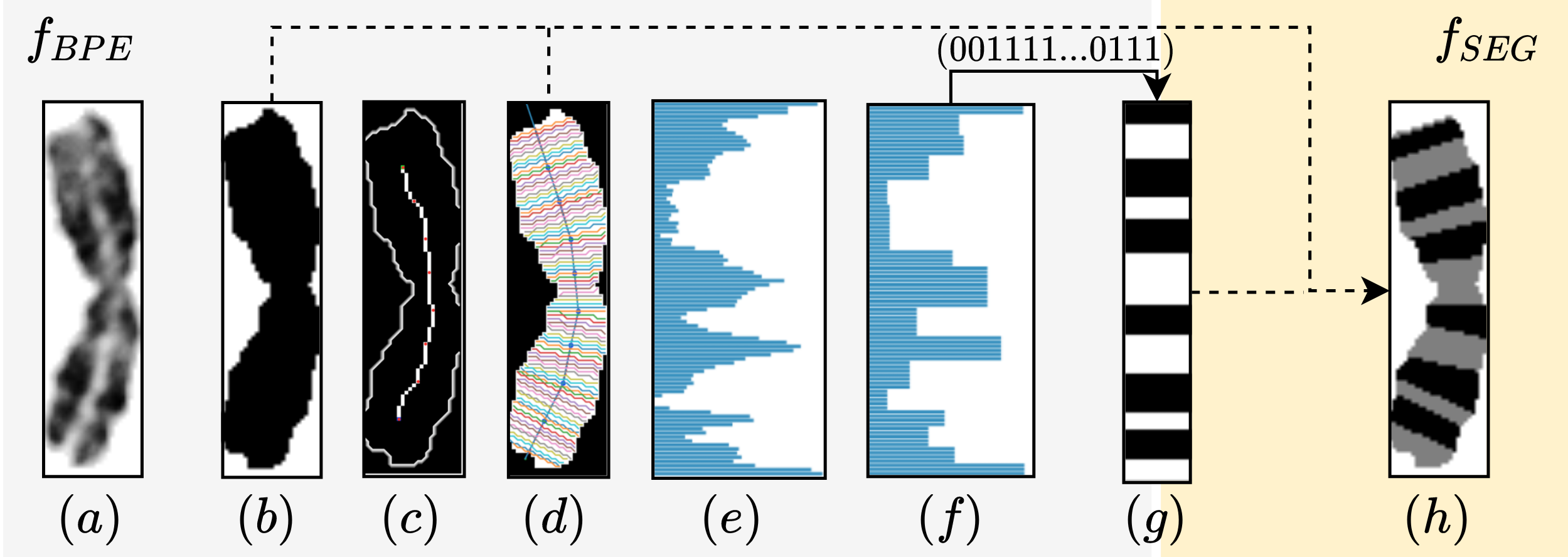}
    \caption{Function $f_{BPE}$ and $f_{SEG}$. From left to right: Chromosome image, shape mask, skeleton, sampling points, density profile, filtered density profile, 1D banding pattern extended to 2D for visualization, banded segmentation mask.}
    \label{fig:bpe_and_seg}
\end{figure}

Let $y_{real} \in \mathbb{R}^{M\times N}$ be a real grayscale chromosome image of spatial dimension $M \times N$ (see Figure \ref{fig:bpe_and_seg}a). The banding pattern extraction function $f_{BPE}$ takes a chromosome image as input, producing a binary banding pattern vector $bp_{real} \in \{0,1\}^K$ such that $f_{BPE}(y_{real})=bp_{real}$, where $K$ denotes the length of the vector (see 2D representation of this vector in Figure \ref{fig:bpe_and_seg}g).




The function $f_{BPE}$ is based on extracting the density profile of a chromosome \cite{granum1982application}, and follows the approach suggested in \cite{wang2008rule} with some modifications. The density profile is extracted by approximating the medial axis of a chromosome shape, $s_{real}$, via several line segments. The shape mask itself is generated through binary segmentation of $y_{real}$ (see Figure \ref{fig:bpe_and_seg}b). At each point $k$ on the medial axis, perpendicular lines are constructed within the shape mask, we denote the rasterized points making up one line as $P_k$ (see Figure \ref{fig:bpe_and_seg}d, perpendicular lines). All grayscale values of $y_{real}$ at $P_k$ are averaged, and make up the density profile value at $k$ (see Figure \ref{fig:bpe_and_seg}e). Unlike \cite{wang2008rule}, we interpolate between angles of these perpendicular lines. This results in less over- and under-sampling of chromosome areas. As in \cite{wang2008rule}, a non-linear filter is applied on the density profile which transforms each band into a uniform density (see Figure \ref{fig:bpe_and_seg}f). Where areas between peaks and valleys correspond to black and white bands respectively. Additionally, we split ambiguous saddle-points into two, assigning each half to the neighboring value, producing the final banding pattern $bp_{real}$ (see Figure \ref{fig:bpe_and_seg}g).

\paragraph{Banded Segmentation Masks Generation.} \label{section:segmentation_masks}
Pix2pix needs a paired set of training data, displaying the same image in different domains. We define the source domain to be banded segmentation masks (see Figure \ref{fig:bpe_and_seg}h), while the target domain is defined as realistic-looking chromosome images. We define a function $f_{SEG}$, it takes as argument a chromosome binary shape mask and banding pattern, such that $f_{SEG}(bp_{real}, s_{real})=\hat{x}_{real}$, where $\hat{x} \in \{0, 127, 255\}^{M\times N}$ is an approximation of the real banded segmentation mask $x$ of the same dimension. The numbers correspond to black bands, white bands, and background respectively. To do so, each value $bp(k)$ is replicated along its corresponding perpendicular points, $P_k$, within the binary shape mask. Some points might not have been sampled during the extraction process in the shape mask. This happens frequently at points located far away from the medial axis, where the chromosome is curving. In these cases, holes are filled with the value of the nearest neighbor. 

\paragraph{Training of pix2pix.} \label{section:pix2pix}
Pix2pix is a conditional GAN (cGAN) that can translate images from a source domain to a target domain. We use it to translate images from banded chromosome segmentation masks (source domain $X$) into realistic-looking chromosomes (target domain $Y$). It is based on a U-Net \cite{ronneberger2015u} type generator $G$, and a patch-based discriminator $D$. Both networks are pitted against each other in a minimax optimization game, $G$ learns how to generate $y \in Y$ with $x \in X$ as input, while $D$ is tasked to decide whether $y$ is synthetic or not. As no real banded segmentation masks exists, we utilize the approximations $f_{SEG}(f_{BPE}(y_{real}), s_{real}) = \hat{x}$. This builds a tuple of paired data $(\hat{x}, y)$ for training, where $\hat{x}$ is translated into $y$ (see Figure \ref{fig:method_overview}a). The learning objective of the network otherwise remains unchanged, and is given in \cite{isola2017image}. After training, $D$ is discarded, as only $G$ is necessary to generate synthetic chromosomes.

\subsection{Abnormal Banding Pattern-Based Chromosome Generation} \label{section:abnormal_chromosomes}
To mimic abnormal chromosomes, we randomly generate banding patterns based on Perlin noise. Considering that real banding patterns consist of clustered black or white regions of various sizes, random noise would be inadequate to mimic the coherence of real banding patterns. Perlin noise has been used, for example, to generate synthetic breast tissues \cite{dustler2015application}. For differentiation purposes, we denote these abnormal banding patterns as $bp_{perlin}$ and the real ones as $bp_{real}$. These Perlin bands can be backprojected to arbitrary chromosome shapes, in the same manner as real ones. This results in abnormal banded segmentation masks which are further employed for generation of chromosome images exhibiting the given band configuration (see Figure \ref{fig:method_overview}b).

\section{Experiment}
\subsection{Materials}
\paragraph{Dataset.} Karyotypes from the PKi-3 data set \cite{ritter2008automatic}, and self-collected karyotypes from various public online resources compose our final G-banding chromosome data set, displaying 400-550 band level resolutions. The former source consists of 612 karyotypes with sizes of $768\times582\times1$ with chromosomes at pro- or metaphase stage. The latter is a compilation of healthy and unhealthy karyotypes, consisting of 445 human karyotype RGB images with varying dimensions and quality.

Single chromosomes are extracted from the karyotypes, resulting in a total of 42684 images. We apply the following operations to all images: (1) the images are transformed to grayscale, (2) square padding is applied to meet the size of the biggest chromosome in the karyotype, (3) the images are resized to 128x128.

We split the chromosome images into a train, validation, and test set with a $0.7$, $0.15$, $0.15$ ratio. Furthermore, we create an additional set of banded segmentation masks, where random Perlin bands are backprojected onto the real chromosome shapes of the test set. This set is titled test\textsubscript{perlin}, contrary to test\textsubscript{real}.



\paragraph{Metrics.} Evaluating the quality of GAN-based synthetic medical images is an ongoing research topic \cite{armanious2020medgan,yi2019generative}. However, the focus of our method lies in conditioning a GAN on banding patterns. Thus, we adopt appropriate metrics.

Firstly, we calculate the dice score between the banded segmentation masks $\hat{x}_{input}$ and $\hat{x}_{fake}$, whereas \textit{input} can be either \textit{real} or \textit{perlin}, omitting the background. As the generator yields $G(\hat{x}_{input}) = y_{fake}$ only, we create the banded segmentation masks for the synthetic chromosome in the same manner as for the real ones. Namely, $\hat{x}_{fake} = f_{SEG}(f_{BPE}(y_{fake}))$ (see Figure \ref{fig:method_overview}b). The dice score measures the area of overlap between two segmentation masks (1 best, 0 worst) and is commonly used in medical segmentation tasks \cite{tajbakhsh2020embracing}. Furthermore, A. S. Pires et al. \cite{pires2016cytogenomic} used dice coefficient to measure similarity of fungi minichromosome banded profiles. We measure dice score on the 2D banded segmentation mask, instead of employing a 1D metric on the banding patterns directly, as it is more robust to variation in the extracted medial axis of $\hat{x}_{fake}$.

We further propose to measure the Mean Absolute Error Number of Bands (MAENB, the lower the better) which measures the mean deviation of bands in the synthetic chromosomes per band compared to the input band:
\begin{equation}
    MAENB = \alpha (||bp_{input}|_{b} - |bp_{fake}|_{b}| + ||bp_{input}|_{w} - |bp_{fake}|_{w}|),
\end{equation}
where $|bp|_{b}$ and $|bp|_{w}$ denote the number of black and white bands in an extracted banding pattern respectively, and $\alpha$ is a normalization factor defined as $\alpha = 1 /(|bp_{input}|_{w} + |bp_{input}|_{b})$. Hence, MAENB can give an insight on the discrepancy between the amount of input and output bands.


\paragraph{Training and implementation details.} 
The implementation is realized in Python, and makes use of the OpenCV\cite{opencv_library} as well as scikit-learn\cite{scikit-learn}. The pix2pix network is taken from the PyTorch implementation by the original authors\footnote{\url{https://github.com/junyanz/pytorch-CycleGAN-and-pix2pix}}, including the hyperparameters, except for the batch size which is set to 32. The network was trained for 100 epochs on a shared cluster with NVIDIA V100 and V100S GPUs on the training set and validated every 10 epochs. Each epoch took roughly 17 minutes to train. By default, the network upscales the images to 256x256, however, for evaluation the images are downscaled to 128x128.



\subsection{Results} 
\paragraph{Quantitative Results.}
Using the metrics, we can define a point of convergence for pix2pix. Table \ref{tab:pix2pix_validation} shows that the network yields the best results around epoch 50 with a dice score of 81.5\% and a MAENB value of 0.111. However, afterward the performance on the validation set decreases. For subsequent analysis, we thus choose the weights obtained after 50 epochs.

\begin{figure}[t]
\centering
  \begin{minipage}[b]{0.39\textwidth}
    \small
    \centering
    \caption{Test set performance (mean and standard deviation).}
    \label{tab:pix2pix_test}
    \vspace{3mm}
    \begin{tabular}{|l|l|l|}
        \hline
         & DICE $\uparrow$ & MAENB $\downarrow$ \\ \hline
        test\textsubscript{real} & \begin{tabular}[c]{@{}l@{}}.815\\$\pm$.101\end{tabular} & \begin{tabular}[c]{@{}l@{}}.109\\ $\pm$.157\end{tabular} \\ \hline
        test\textsubscript{perlin} & \begin{tabular}[c]{@{}l@{}}.710\\ $\pm$.100\end{tabular} & \begin{tabular}[c]{@{}l@{}}.162\\ $\pm$.139\end{tabular} \\ \hline
        baseline & \begin{tabular}[c]{@{}l@{}}.496\\ $\pm$.113\end{tabular} & \begin{tabular}[c]{@{}l@{}}.304\\ $\pm$.271\end{tabular} \\ \hline 
    \end{tabular}
    
  \end{minipage}
  \hfill
  \begin{minipage}[]{0.6\textwidth}
    \centering
    \includegraphics[width=0.9\textwidth]{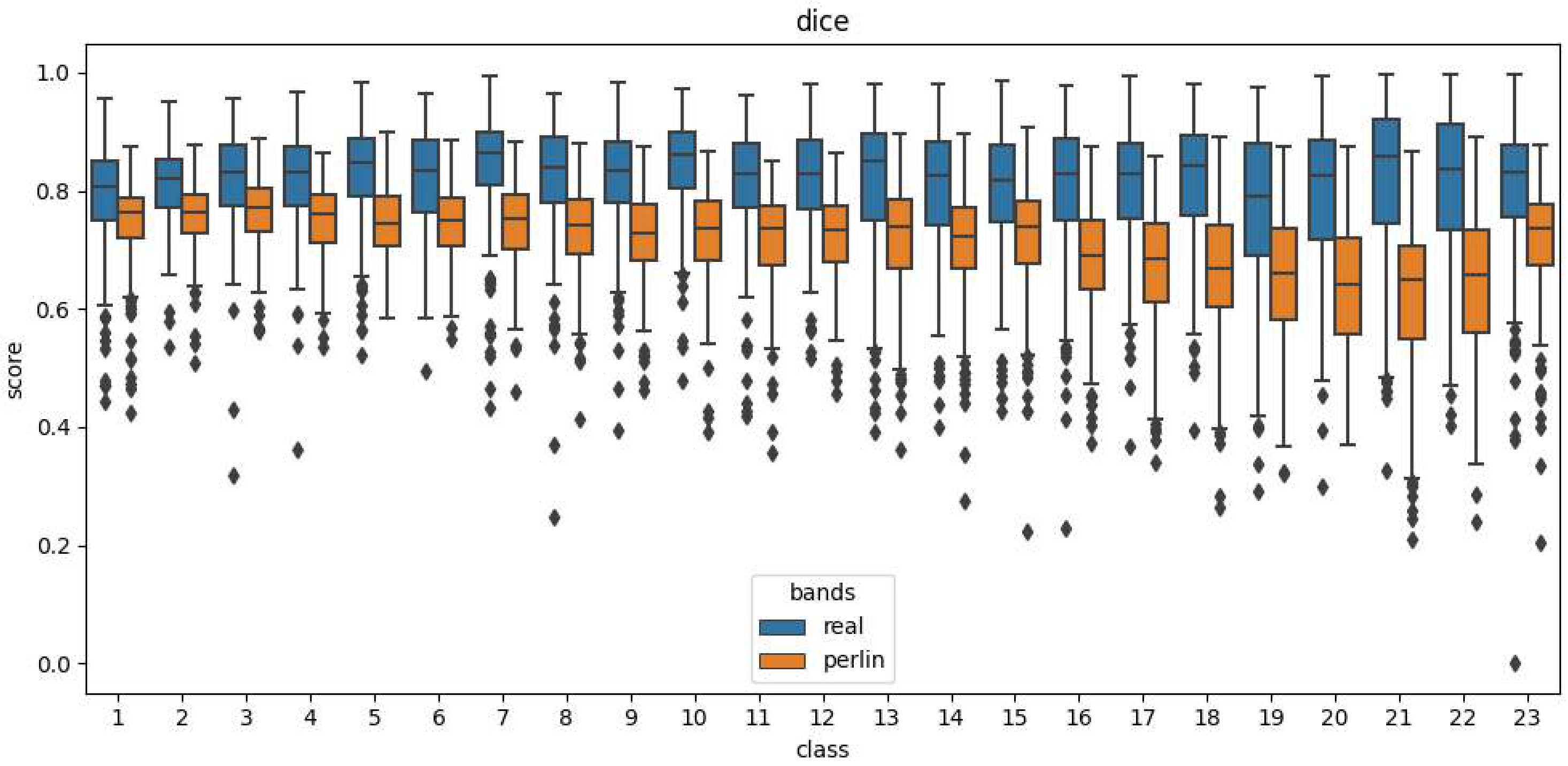}
    \caption{Test Dice Score Per Class}
    \label{fig:test_per_class}
      
    \end{minipage}
\end{figure}

Final test results are nearly identical for the real bands compared to the validation set (see Table \ref{tab:pix2pix_test}). Nonetheless, a decline in performance for the test\textsubscript{perlin} set can be observed by roughly 10\% in dice score and an increase of MAENB by around 0.05. We further compare the corresponding $\hat{x}^i_{perlin} \in$ test\textsubscript{perlin} with $\hat{x}^i_{real} \in$ test\textsubscript{real} for all $i$ as a baseline, where $i$ denotes the sample number. Consider that these corresponding masks have the same shape but only differ in their banding patterns. Doing so reveals that the real and Perlin bands are uncorrelated, being that the final dice score is around random binary guessing probability. It further serves as a baseline as it demonstrates the use-case where no conditioning of banding patterns, but the shape only, is possible. This shows that the generator is able to partially impose the abnormal banding patterns, as test\textsubscript{perlin} achieves a dice score of 71\%, significantly higher than the baseline.

Other state-of-the-art methods employing dice score as a measure of similarity for a segmentation task involving cGANs for image-translation report similar scores. For example, C. Chen et al. \cite{chen2020realistic} obtained a max. averaged dice score of 83\% on a semi-supervised segmentation task. In addition, Yan et al. \cite{yan2019domain} used CycleGAN for domain adaption of MR images, trained a U-Net and achieved dice score for two data sets of 80.5\% and 86.7\% respectively. 


The dice score is relatively consistent across chromosome classes, with the exception of higher classes with Perlin bands as input (see Figure \ref{fig:test_per_class}). Chromosomes 1 to 22 are assigned their numbers based on size, apart from chromosome 22 which is slightly larger than 21 \cite{gersen1999principles}. Hence, the drop in performance might be due to less variability in expressible banding patterns of short length. Similarly, the performance on class 23 likely might increases again, as we do not differentiate between the longer X and the shorter Y chromosome during evaluation.


\begin{table}[t]
\caption{Validation set performance every 10 epochs (mean and standard deviation).}
\label{tab:pix2pix_validation}
\centering
\small
\begin{tabular}{|l|l|l|l|l|l|l|l|l|l|l|}
\hline
& 10 & 20 & 30 & 40 & 50 & 60 & 70 & 80 & 90 & 100 \\ \hline
DICE $\uparrow$ & \begin{tabular}[c]{@{}l@{}}.765\\ $\pm$.107\end{tabular} & \begin{tabular}[c]{@{}l@{}}.783\\ $\pm$.111\end{tabular} & \begin{tabular}[c]{@{}l@{}}.799\\ $\pm$.109\end{tabular} & \begin{tabular}[c]{@{}l@{}}.814\\ $\pm$.104\end{tabular} & \textbf{\begin{tabular}[c]{@{}l@{}}.815\\ $\pm$.102\end{tabular}} & \begin{tabular}[c]{@{}l@{}}.805\\ $\pm$.107\end{tabular} & \begin{tabular}[c]{@{}l@{}}.806\\ $\pm$.105\end{tabular} & \begin{tabular}[c]{@{}l@{}}.793\\ $\pm$.107\end{tabular} & \begin{tabular}[c]{@{}l@{}}.778\\ $\pm$.111\end{tabular} & \begin{tabular}[c]{@{}l@{}}.782\\ $\pm$.111\end{tabular} \\ \hline
MAENB $\downarrow$& \begin{tabular}[c]{@{}l@{}}.154\\ $\pm$.186\end{tabular} & \begin{tabular}[c]{@{}l@{}}.139\\ $\pm$.186\end{tabular} & \begin{tabular}[c]{@{}l@{}}.120\\ $\pm$.171\end{tabular} & \textbf{\begin{tabular}[c]{@{}l@{}}.111\\ $\pm$.155\end{tabular}} & \begin{tabular}[c]{@{}l@{}}.111\\ $\pm$.162\end{tabular} & \begin{tabular}[c]{@{}l@{}}.116\\ $\pm$.165\end{tabular} & \begin{tabular}[c]{@{}l@{}}.113\\ $\pm$.160\end{tabular} & \begin{tabular}[c]{@{}l@{}}.128\\ $\pm$.171\end{tabular} & \begin{tabular}[c]{@{}l@{}}.140\\ $\pm$.180\end{tabular} & \begin{tabular}[c]{@{}l@{}}.135\\ $\pm$.174\end{tabular} \\ \hline
\end{tabular}

\end{table}

\paragraph{Qualitative Results.}

\begin{figure}[t]
    \centering
    \includegraphics[width=0.86\textwidth]{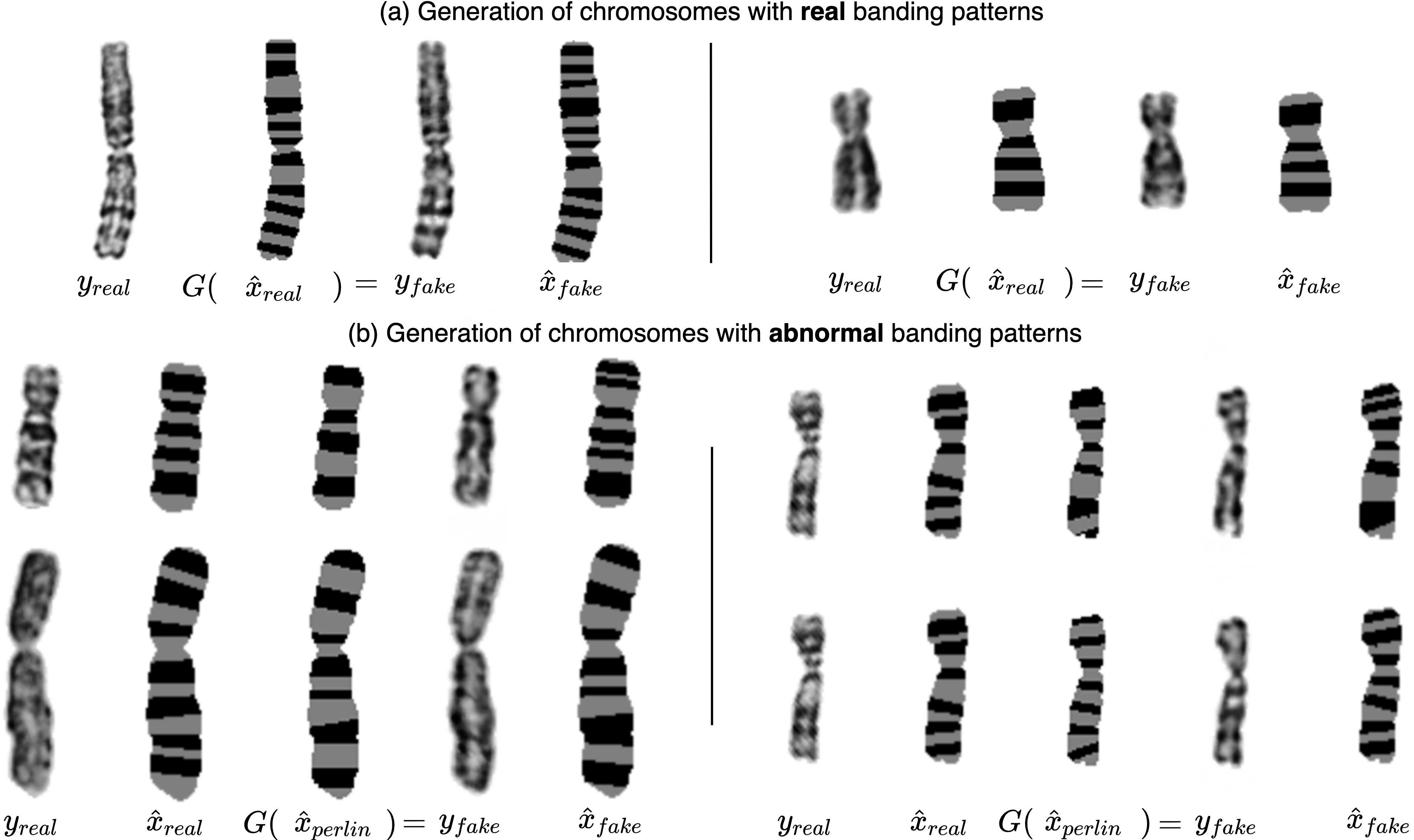}
    \caption{Chromosome generations with $\hat{x}_{real}$ (a) and $\hat{x}_{perlin}$ (b) as input. The generator outputs $y_{fake}$, the banded segmentation mask $\hat{x}_{fake}$ of the output visualizes how well the banding patterns are imposed by the generator. In addition, $y_{real}$ and $\hat{x}_{real}$ are given for comparisons sake. For naming conventions see Figure \ref{fig:method_overview}.}
    \label{fig:generations_diagram}
\end{figure}

Figure \ref{fig:generations_diagram}a shows images $y_{fake}$ which are generated by pix2pix with real banding patterns as input. The quality of the synthetic chromosomes is similar to the real ones ($y_{real}$). In most of the cases, $\hat{x}_{real}$ and $\hat{x}_{fake}$ look alike, however, some bands are split in two or merged into a single one. This might also be a fault of the hand-crafted nature of the banding pattern extraction itself. Further comparing $\hat{x}_{real}$ with $y_{fake}$, reveals that the synthetic chromosomes indeed display the desired input banding patterns.

In the same manner, Figure \ref{fig:generations_diagram}b shows synthetic abnormal chromosomes generated with $\hat{x}_{perlin}$ as input. Overall, the generation quality seems to be on a similar level compared to real bands. In some cases, $\hat{x}_{perlin}$ and $\hat{x}_{fake}$ are very similar and only slightly differ in the size of the bands (see Figure \ref{fig:generations_diagram}b top left). In other cases, some of the bands are correctly positioned and have the correct size, but are missing or adding some bands. This can be seen in Figure \ref{fig:generations_diagram}b bottom left row, where $\hat{x}_{fake}$ resembles a mix between $\hat{x}_{perlin}$ and $\hat{x}_{real}$ which suggest that the generator partially overfits on chromosomes from the training set.

We further analyze how the generator behaves on banded segmentation masks with a shape mask but varying banding patterns (two rows on the right of Figure \ref{fig:generations_diagram}b). Even though the same shape is used, the bands are clearly distinct in between each other and mostly exhibit the input banding patterns. Overall, the results suggest that this approach can be utilized to generate abnormal chromosomes with more fine-grained control over the exhibited banding patterns. However, the input banding pattern is not always completely imposed.


\section{Conclusion} 

In this work, we propose a method of conditionally generating chromosome images based on banding patterns. Our method can be applied to chromosome data sets without annotations, as banded segmentation masks are created in an automated fashion. We validate on a test set of 6432 single chromosome images while using real as well as abnormal patterns. Synthetic chromosomes are of high visual quality when conditioning on real as well fake banding patterns up to a promising extent. However, banding patterns are not always strictly imposed which could be addressed through domain specific adaption of the learning objective in future work.
We believe that this approach can be exploited for data augmentation purposes of healthy and unhealthy chromosomes, displaying deviations from the standardized ideograms such as the recombination of sections, or missing bands, improving on existing image-based methods in the field of cytogenetics.
 
\section*{Acknowledgments}
We thank András Kozma and Robert Caldwell for their feedback.

\bibliographystyle{unsrt}  
\bibliography{references}

\end{document}